\documentclass[12pt]{article}

\usepackage{color}

\usepackage{amsmath}
\usepackage[authoryear]{natbib}
\usepackage{latexsym, epsfig, amssymb, amsmath, amsthm, graphicx, mathrsfs,multirow}
\usepackage[OT1]{fontenc}
\usepackage{anysize}
\usepackage{subfigure}
\usepackage{tabularx}
\usepackage{pdflscape}
\usepackage{algorithm}
\usepackage{algorithmic}
\usepackage{blindtext}
\usepackage{enumitem}
\usepackage{adjustbox}

\def\singlespace{\def\baselinestretch{1}\@normalsize}
\def\vm{1.28}
\def\vt{1.28}
\renewcommand{\baselinestretch}{\vt}
\marginsize{1.2in}{0.9in}{0.5in}{1.3in} %
\makeatletter

\usepackage[compact]{titlesec}
\usepackage{booktabs}

\newtheorem{lemma}{Lemma}
\newtheorem{proposition}{Proposition}
\newtheorem{theorem}{Theorem}
\newtheorem{condition}{Condition}

\makeatother

\newcommand{\bSigma}{\mbox{\boldmath $\Sigma$}}

\newcommand{\EE}{\mathbb{E}}
\newcommand{\PP}{\mathbb{P}}

\newcommand{\RR}{\mathbb{R}}

\newcommand{\Cov}{\mathrm{Cov}}

\newcommand{\tr}{\mathrm{tr}}

\newcommand{\vecc}{\mathrm{vec}}

\newcommand{\supp}{\mathrm{supp}}

\newcommand{\cL}{\mathcal{L}}
\newcommand{\cI}{\mathcal{I}}

\newcommand{\XX}{\mathbb{X}}

\def\T{{ \mathrm{\scriptscriptstyle T} }}

\DeclareMathOperator*{\argmin}{arg\,min}

\begin{document}

\DeclareGraphicsExtensions{.pdf,.gif,.jpg,.ps,.eps}

\title{High-Dimensional Interactions Detection with Sparse Principal Hessian Matrix}

\date{}
\author{Cheng Yong Tang, Ethan X. Fang, and Yuexiao Dong}

\maketitle

\renewcommand{\baselinestretch}{\vm}
\normalsize

\begin{abstract}

In statistical learning framework with regressions, interactions are the contributions to the response variable from the products of the explanatory variables. In high-dimensional problems, detecting interactions is challenging due to  combinatorial complexity and limited data information. We consider detecting interactions by exploring their connections with the principal Hessian matrix. Specifically, we propose a one-step synthetic approach for estimating the principal Hessian matrix by a penalized M-estimator. An alternating direction method of multipliers (ADMM) is proposed to efficiently solve the encountered regularized optimization problem. Based on the sparse estimator, we  detect the interactions by identifying its nonzero components. Our method directly targets at the interactions, and it requires no structural assumption on the hierarchy of the interactions effects.  We show that our estimator is theoretically valid, computationally efficient, and practically useful for detecting the interactions in a broad spectrum of scenarios.

\end{abstract}


\renewcommand{\baselinestretch}{\vm}
\normalsize

\newpage

\section{Introduction}

The interaction effect is an important consideration in regression problems that are commonly encountered in practice.  It refers to a general situation when it is not adequate to build a model with the original explanatory variables alone in a simple additive way -- referred to as the main effects. At the same time, it could be more effective to build a model incorporating constructed variables from the  products of the variables -- referred to as the interactions between the variables; see, for example, the monograph of \cite{JamesJ.Jaccard2003} for an overview. Since the interactions are constructed from the explanatory variables, it is clear that  they can be  further expanded in a hierarchical manner. That is, the construction can continue searching for  higher order  interactions  that can be defined based on all variables including those previously constructed ones.
In current paradigm of scientific investigations such as genome-wide association
studies (GWAS), attempting to find interactions has received substantial recent attentions; see, for example,  \cite{Cordell2009,Dato2018,Ritchie2018}. 
Indeed,   
as pointed out in \cite{Ritchie2018}, the development of  modeling approaches and computational techniques for identifying interaction items will enable practitioners to effectively identify important genetic features for new advances in medical studies.

Recently, there has been a surge in the development of methods for solving high-dimensional regression problems.  For an overview of the current state of knowledge, we refer to the monographs  \cite{Buhlmannvan_2011},   \cite{Hastieetal_2015},  the review by  \cite{FanLv_2009_Sinica} and  references therein.
Conceptually speaking,  existing penalized regression methods can be applied by incorporating the constructed interactions as new variables. However,  because of the aforementioned interactive and hierarchical nature, the complexity of the problem scales up very quickly, both computationally for practical implementations and theoretically for analyzing the properties of the estimators.  For example, in a typical situation with $p$ explanatory variables, the number of the possible two-way interactions is $p(p-1)/2$, growing at a quadratic rate of~$p$ that itself is typically assumed growing at some exponential rate of the sample size~$n$.   Consequently,   on one hand,  with the total number of explanatory variables including interactions at the order of $\mathcal{O}(p^2)$,  the computational complexity increases substantially for estimation.  On the other hand, due to the increased data dimensionality, the relative signal strength from the explanatory variables becomes substantially weaker, and makes detection these signals much more challenging.
 
There also has been active development on methods for detecting and estimating the interactions in high-dimensional regression problems. 
To address the challenges,  a main strategy is to impose some hierarchical structure between the main effect  and the interactions. That is, in the absence of the original variables, the interactions alone do not contribute to the model.  This is referred to as the marginality principle \citep{Nelder1977}. 
In  a class of the existing methods guided by this principle, the interactions are detected and estimated by some two-step or iterative two-step procedures that first select the main effects by some screening or penalized regression methods, and then perform a second run of selections by incorporating all potential interactions from the selected variables.  For examples of the methods belonging to this class, we refer to \cite{Yuan2009}, \cite{Zhao2009}, \cite{Choi2010}, \cite{Bien2013}, \cite{Hao2014},  \cite{Shah2016},  \cite{Haris2016}, and \cite{hao2018model}. 
More recently, \cite{She2018} investigate the problem by a group penalty approach incorporating the structural hierarchy. 
Requiring no hierarchical structure and in a setting with multiple response, \cite{Kong2017a} propose to search for interactions by a sure screening procedure based on the square functions of the distance correlation measures.  This approach is computationally efficient, and only requires ranking with pairwise quantities.  From the inverse modeling perspective, 
\cite{Jiang2014} consider testing procedures by the sliced inversion regression  approach \citep{Li_JASA_1991} for variable selections of the first- and second-order effects. For classification problems and general index models, \cite{Li2017b} study some information criteria based approaches for variable and interaction selections.  
Recently, \cite{Fan2016} propose a method to detect the signals from the interactions by examining the pairwise covariances between the squared response and each explanatory variable.  Upon evaluating the covariances, \cite{Fan2016}  propose a new two-step procedure that first conducts sure screening for the interactions, and then conducts a second step of penalized regression without imposing the hierarchical~structure.

For those two-step procedures,
detecting interactions in the second step upon selecting some variables such as the main effect has a clear limitation in the high-dimensional scenarios.  Particularly for the ones with some hierarchical structures, if the signal strength relative to the noise from the main effects is weak, then the method in the first step may miss the main effect. Subsequently, it becomes impossible to detect the interactions in the second step.  Under  high-dimensional settings,  the noise aggregates as the dimension increases, and decreases the relative signal strength. Such weaker  signals make missing main effects in the first step more likely to happen.

In this paper, we propose a new method to detect the interaction effects in regression problems by  a one-step penalized M-estimator.
Our method does not assume a  hierarchical structure, and it also requires  no screening step.
   Our method is developed by utilizing the so-called principal Hessian matrix \citep{Li1992}, which is defined as the expectation of the second order derivative of the mean function.  To solve the challenging high-dimensional problem,  we assume that the principal Hessian matrix is sparse, reflecting the reality that given limited data information, only few meaningful interactions can be supported with desired estimation accuracy.
A sparsity encouraging M-estimator is proposed by minimizing a dedicated crafted penalized squared loss function.  As shown in our development, this procedure involves a matrix optimization problem, which is substantially more challenging than those vector optimization problems \citep{ding2014introduction}.   
To meet the challenge, we propose an alternating direction method of multipliers (ADMM) based algorithm to compute the estimator efficiently.
We show  by simulations that the proposed method outperforms existing methods, and our theory confirms that the estimator works satisfactorily, allowing the dimensionality of the explanatory variables growing exponentially with the sample size.

The rest of this paper is organized as follows. The proposed method using the principal Hessian matrix  for detecting interactions is outlined in Section \ref{s2}, followed by the  alternating direction method of multipliers (ADMM) for solving the optimizations  elaborated in Section \ref{s3}.  Numerical examples including simulation and a real data example are presented in Section \ref{s:sim}  to demonstrate the promising performance of the method. Section \ref{s5} gives  theoretical analysis, and the proofs are provided in the Appendix.

%
%
%

\section{Methods} \label{s2}
\subsection{Principal Hessian Matrix}

The key device in our method is the principal Hessian matrix.  In \cite{Li1992}, the principal Hessian matrix is proposed  as a powerful device for investigating dimension reduction and data visualization. 
We consider the regression problem with response variable $Y$  and explanatory variable $X=(X_1,\dots,X_p)^\T \in {\mathbb R}^p$ that is taken as a random vector.  Let $m(X)=\EE(Y|X)$ be the conditional mean function.  
The principal Hessian matrix of \cite{Li1992} is  then defined as $\Psi=\EE\{\partial^2 m(X)/\partial X\partial X^\T\}$ where the expectation is taken with respect to the joint distribution of $(Y,X^\T)^\T$.   
By  the Stein's Lemma \citep{Stein1981},  \cite{Li1992} shows that if $X\sim N(0,\Sigma_X)$, then the  principal Hessian matrix $\Psi$ satisfies
\begin{align}\label{eq:stein}
\Psi= \Sigma^{-1}_X \Sigma_{YXX}\Sigma^{-1}_X, 
\end{align}
where  $ \Sigma_{YXX}=\EE[\{Y-\EE(Y)\}XX^\T]$.   The principal Hessian matrix is an effective device for the so-called sufficient dimension reduction; see \cite{Li1992} and \cite{Cook1998}.

Since  $m(X)=\EE(Y|X)$,  a model for $Y$ can be written as $Y=m(X)+\varepsilon$ with $\varepsilon$ being a zero mean random variable independent of $X$.  
For ease of presentation and without loss of generality, we assume that $X$ is centered so that all of its components $X_1,\dots,X_p$ are of  mean zero. 
We now demonstrate the connection between the principal Hessian matrix $\Psi$  in (\ref{eq:stein}) and the interaction effects.  Let us consider the following working model with both the main and interaction effects:
\begin{align}\label{eq:model}
Y=m(X)=\sum_{i=1}^p \theta_i X_i+\sum_{i=1}^p \sum_{j=i}^p \beta_{ij}X_i X_j+ \varepsilon. 
\end{align}
Then,  by (\ref{eq:model}), it holds that
\begin{align}\label{eq:inter}
\EE\left\{ \frac{\partial^2 m(X)}{\partial X_i \partial X_j} \right\}=
\begin{cases}
\psi_{ij}=\psi_{ji}=\beta_{ij}, \mbox &\text{when }i\neq j,\\
2\psi_{jj} = 2\beta_{jj},\mbox & \text{when } i=j,
\end{cases}
\end{align}
where $\psi_{ij}$ is the $(i,j)$-th component of $\Psi$. 
For the case $i\neq j$,  the first equality in (\ref{eq:inter}) is by  definition~(\ref{eq:stein}),  the second equality is due to that $\Psi$ is symmetric, and the third equality is from the working model (\ref{eq:model}). 
Hence, it is intuitive that  the principal Hessian matrix $\Psi$ is a very informative device for detecting interactions.   
Immediately from (\ref{eq:stein}) and (\ref{eq:inter}), we have the following result for detecting interactions with $\Psi$. 
\begin{proposition}\label{prop 1}
If  $X\sim N(0, \Sigma_X)$,  then 
under model (\ref{eq:model}), $\psi_{ij}=\psi_{ji}\neq 0$ if and only if $\beta_{ij}\neq 0$.
\end{proposition}
The normality assumption in Proposition \ref{prop 1} is inherited from the Stein's Lemma. 
For detecting interactions in model (\ref{eq:model}), the normality assumption  can be relaxed. 
%
For example, in the ideal case that $\Sigma_X=I$, then $\Psi=\Sigma_{YXX}$, and $\psi_{ij}=\psi_{ji}=\beta_{ij}$ remains true provided that $\EE(X_i X_j X_k)=0$ $(i,j,k=1,\dots,p)$.  
%
%
%
%
%
For a general $\Sigma_X$, the normality  assumption in Proposition \ref{prop 1} can be replaced by  the following  moment conditions: 

\begin{condition}\label{con:0}
$X=(X_1,\dots,X_p)^\T$ satisfies
\begin{enumerate}[align=left]

\item[C1.]  $\EE(X_i X_j X_k)=0$ for any $i,j,k=1,\ldots, p$;

\item[C2.]   Let $Z=(Z_1,\dots,Z_p)^\T=\Sigma_X^{-1}X$. It holds that $\Cov(X_kX_\ell, Z_iZ_j)\neq 0$ if and only if $k=i,\ell=j$ or $k=j,\ell=i$.  
\end{enumerate}
\end{condition}

\begin{proposition}\label{prop:2}
Under Condition \ref{con:0} and model (\ref{eq:model}), $\psi_{ij}=\psi_{ji}\neq 0$ if and only if $\beta_{ij}\neq 0$.
\end{proposition}

%
%
%
%
%
%
%
%
%
%
%
%

The proof of Proposition \ref{prop:2} is given in the Appendix. 
Condition \ref{con:0} is clearly valid for the normal distribution, and it holds more broadly.   For example, it holds for the elliptical symmetric distributions whose characteristic function takes the form $\phi(t^\T \Sigma_X t)$. Other practically appealing situations for  Condition \ref{con:0} include when $X$ contains independent components, and/or there are only few nonzero $\beta_{ij}$ in (\ref{eq:model}). 

Here C1 of Condition \ref{con:0}  ensures that the linear contribution from $X$ in $m(X)$ does not interfere with that from the interactions.
We further note that if there is no linear contribution from $X$, or the main effect  has been consistently estimated and removed from $m(X)$,   then C1 in Condition \ref{con:0}  is not required for detecting the interactions.  That is,  consistent estimation of the  main effects  helps alleviating the conditions on the distribution of $X$.  Existing high-dimensional penalized regression methods, e.g., those  in  \cite{FanLv_2009_Sinica},  \cite{Buhlmannvan_2011}, and  \cite{Hastieetal_2015} can potentially be performed for such a purpose, but some cautions are needed.  To ensure that the linear effects can be adequately recovered in a working model with no interactions, some conditions are required.   A sufficient condition is that $\EE\{X_i (X_jX_k)\}=0$ for all $i$ such that $\theta_i\neq 0$, and all $(j,k)$ such that $\beta_{jk}\neq 0$.  Furthermore, 
C2 of Condition \ref{con:0} is on the correlations between  $X_{k}X_\ell$ and $Z_{i}Z_j$. 
 Since $\Cov(Z,X)=\Cov(\Sigma_X^{-1}X,X)=I$,  zero correlations  between $X_k$ and $Z_i$ for $k\neq i$ are already ensured.  Instead of imposing the normality assumption, C2 in Condition  \ref{con:0} is seen as on correlations between second order terms, which can be verified empirically; and transformation may be developed if concerns arise.


We note that by its definition, $\Psi$ is suitable for studying interactions between numerical explanatory variables. For categorical variables,  dummy variable coding is needed, resulting in sub-groups of the observations so that our methods and other penalized regression analysis can be equally applied within the resulting groups for detecting the interactions between numerical and categorical variables.




\subsection{Sparse Estimation of $\Psi$}

Our study intends to explore an effective estimator for $\Psi$ under  high-dimensional settings, and then to detect the interactions. 
In high-dimensional regression problems, sparsity is a useful notion for statistical inferences; see, among others, \cite{Hastieetal_2015}. In the context of interactions in $m(X)$, it means that the total number of contributing pairs of the explanatory variables is small, so that $\Psi$ is sparse.  Thus, exploring the interactions becomes a sparse matrix estimation problem.

By examining $\Psi$ in (\ref{eq:stein}), we observe that a sparse estimation of this $p\times p$ matrix is challenging; and  plugging-in individual components is not feasible for solving high-dimensional problems.  First, estimating $\Sigma_X^{-1}$ is difficult when $p$ is large. Though existing methods can be applied, e.g., those reviewed  in \cite{Fan2016}, there are two major concerns. First is on whether or not it is suitable to impose assumptions such as $\Sigma_X^{-1}$ being sparse. Second is that  if one constructs an estimator of $\Psi$ from sparse estimators of $\Sigma_X^{-1}$ and~$\Sigma_{YXX}$, the implied sparse components of $\Psi$ may not be the desirable ones in~$\Psi$.
As an example,  we consider $Y=X_1+X_1 X_2 +\epsilon$, $X=(X_1,X_2,X_3)^\T$, $X\sim N(0, \Sigma_X)$ with $\Sigma_X=(\sigma_{ij})_{i,j=1}^3$,  and $\sigma_{ij}=\rho^{|i-j|}$ $(i,j=1,2,3)$. Then we have
\begin{align*}
\Sigma_X^{-1}=
\frac{1}{1-\rho^2}
\begin{pmatrix} 
 1 & -\rho &  0 \\
 -\rho & 1+\rho^2 & -\rho\\
 0 &  -\rho & 1
   \end{pmatrix}
   \mbox{ and }
 \Sigma_{YXX}=  \begin{pmatrix}
 2\rho & 1+\rho^2 &  \rho+\rho^3 \\
 1+\rho^2 &  2\rho & 2\rho^2\\
 \rho+\rho^3 &  2\rho^2 & 2\rho^3
   \end{pmatrix}.   
   \end{align*}
  Then,  some algebra shows that in this case the principal Hessian matrix is
   $$\Psi=\begin{pmatrix}
 0 & 1 &  0 \\
 1 & 0 & 0\\
 0 &  0 & 0
   \end{pmatrix}.$$
That is,  $\Psi$ precisely reflects the signals from the interactions, despite that both $\Sigma_X^{-1}$ and $\Sigma_{YXX}$ are  dense. 

In our investigation, we consider the case that $\Psi$ is sparse, with no  sparsity requirements on $\Sigma_X$, $\Sigma_X^{-1}$, and $\Sigma_{YXX}. $  All our technical development is based on assumptions directly on $\Psi$. Specifically, we aim at a one-step sparse estimator of $\Psi$.   The key development is the following. First, we observe that from (\ref{eq:stein}), $\Sigma_X\Psi\Sigma_X=\Sigma_{YXX}$. This motivates to consider the weighted quadratic  loss
\begin{align}
\mathcal{L}(\Psi)=\tr\{ \Sigma_X^{-1}(\Sigma_X\Psi\Sigma_X-\Sigma_{YXX})^\T\Sigma_X^{-1}(\Sigma_X\Psi\Sigma_X-\Sigma_{YXX})\},
\end{align}
which is minimized at $\Psi = \Sigma^{-1}_X \Sigma_{YXX}\Sigma^{-1}_X$.  With some elementary algebra, it can be shown that $\mathcal{L}(\Psi)$ depends on $\Psi$ only through
\[
\tr(\Psi^\T \Sigma_{X}\Psi \Sigma_X)-2 \tr(\Psi \Sigma_{YXX}).
\]
We denote by $\{y_i, x_i\}_{i=1}^n$ the observed data.
Then, we propose to replace $\Sigma_X$ and $\Sigma_{YXX}$ by their sample counterparts: $S=n^{-1}\sum_{i=1}^n x_ix_i^\T$,  $Q=n^{-1}\sum_{i=1}^n (y_i -\bar y) x_ix_i^\T$ with $\bar y=n^{-1}\sum_{i=1}^n y_i$, and consider sparse inducing $\ell_1$-penalized estimation.  Concretely,  we propose an  M-estimator:  
\begin{align}\label{eq:est}
\widehat\Psi=\argmin_{\Psi}\Big\{  \tr(\Psi^\T S \Psi S)/2  - \tr(\Psi Q)+\lambda\|\Psi\|_1\Big\},
\end{align}
where $\|A\|_1$ is the $\ell_1$-norm of the matrix $A$, and $\lambda>0$ is a tuning parameter.
Given $\widehat \Psi = (\widehat\psi_{ij})_{p\times p}$, we propose to detect interactions as 
\begin{align}\label{set}
{\cal M}=\{(i,j): \widehat\psi_{ij}\neq 0,  i\leq j \}. 
\end{align}
Here the function in (\ref{eq:est}) is quadratic in the matrix $\Psi$, which is different  from those penalized estimators for handling vectors in the literature so that existing algorithms for solving penalized regressions do not apply.  Because only linear operators are involved in (\ref{eq:est})  with no inverse of a large matrix,  the calculation of the loss is computationally efficient and scalable. In the next section, we design an efficient algorithm to compute~$\widehat\Psi$.  
We also note that  $\widehat\Psi$ may not symmetric. For practical applications, we suggest symmetrizing it by $(\widehat\Psi+\widehat\Psi^\T)/2$. 
 
\section{Algorithm} \label{s3}

\subsection{Alternating Direction Method of Multipliers}

We first observe that  the Hessian of  problem \eqref{eq:est} is $S\otimes S$ and positive-semidefinite, where $\otimes$ denotes the Kronecker product.
The loss function in (\ref{eq:est}) is thus  convex.
We propose to solve (\ref{eq:est}) by the alternating direction method of multipliers (ADMM). Our algorithm is inspired by the algorithm developed in \cite{Jiang2015}, in which a large matrix for classification with quadratic discriminant analysis is directly estimated in the same spirit of (\ref{eq:est}).


ADMM  was first proposed in \cite{glowinski1975approximation}, which is essentially a splitting version of the augmented Lagrangian method to solve optimization  problems with a separable objective under a linear constraint:
$$
\min_{\beta,\eta\in\RR^p} g_1(\beta) + g_2(\eta), \text{ subject to }A\beta + B\eta = b, 
$$
where $A,B\in\RR^{k\times p}$, $b\in\RR^p$, and $g_1,g_2:\RR^{p}\rightarrow \RR$ are continuous functions. Recently, ADMM finds wide applications in different fields such as statistics, image processing and machine learning. This is due to the algorithm's easy implementation and practical efficiency;  see \cite{ADMMreview} and \cite{eckstein2012augmented} for some  reviews.

To apply the ADMM algorithm to compute our estimator,  
we rewrite  problem \eqref{eq:est} into the following equivalent form to facilitate the algorithm design that
\begin{equation}\label{eqn:adm}
\min_{\Psi,\Phi} \tr(\Psi^\T S\Psi S)/2 - \tr(\Psi Q) + \lambda\|\Phi\|_1, \text{ subject to }\Psi-\Phi = 0.
\end{equation}
The augmented Lagrangian dual problem associated with the above problem is 
$$
\cL_\rho(\Psi,\Phi,\Lambda) = \tr(\Psi^\T S\Psi S)/2 - \tr(\Psi Q) + \lambda\|\Phi\|_1 + \langle \Lambda,\Psi-\Phi\rangle + \frac{\rho}{2} \|\Psi-\Phi\|_F^2,
$$
where $\Lambda$ is the dual variable associated with the equality constraint, and $\rho>0$ is a penalty parameter. The ADMM algorithm runs iteratively, at the $(t+1)$-th iteration, we update the solutions by
\begin{equation}\label{eqn:admm}
\begin{aligned}
\Psi^{t+1}  & = \argmin_\Psi \cL_\rho (\Psi,\Phi^t,\Lambda^t),\\
\Phi^{t+1} & = \argmin_\Phi \cL_\rho(\Psi^{t+1},\Phi,\Lambda^{t}),\\
\Lambda^{t+1} & = \Lambda^t + \rho(\Psi^{t+1} - \Phi^{t+1}).
\end{aligned}
\end{equation}
Note that since our problem is convex, and the objective value is lower bounded, the convergence result of the algorithm has been well established in existing literature;  see \cite{fang2015generalized} and \cite{sun2015convergent} for examples.
 We further point out that  in our simulation studies in Section \ref{s:sim}, the algorithm performs satisfactorily empirically.

\subsection{Solving the Subproblems}
The key  to implement the ADMM  algorithm developed above is to solve the $\Psi$- and $\Phi$-subproblems in \eqref{eqn:admm} efficiently.  In this subsection, we derive efficient solutions for the two subproblems. For the $\Psi$-subproblem, we have that
$$
\begin{aligned}
\Psi^{t+1}  = & \argmin_{\Psi}\tr(\Psi^\T S\Psi S)/2 - \tr(\Psi Q) + \tr\langle \Lambda^t,\Psi\rangle + \frac{\rho}{2} \|\Psi-\Phi^t\|_F^2 \\
 = & \argmin_{\Psi}  \text{vec}(S\Psi)^\T\text{vec}(S\Psi) /2 - \text{vec}(Q)^\T\text{vec}(\Psi) + \text{vec}(\Lambda^t)^\T\text{vec}(\Psi) + \frac{\rho}{2} \| \text{vec}(\Psi) - \text{vec}(\Phi^t)\|_2^2 \\
 = & \argmin_{\Psi}   \text{vec}(\Psi)^\T\big\{(S\otimes I_p)^\T(S\otimes I_p)\big\}\text{vec}(\Psi) /2 - \text{vec}(Q)^\T\text{vec}(\Psi) + \text{vec}(\Lambda^t)^\T\text{vec}(\Psi) \\
  & + \frac{\rho}{2} \| \text{vec}(\Psi) - \text{vec}(\Phi^t)\|_2^2,
\end{aligned}
$$
where $\otimes$ denotes Kroneker product. Solving this problem is computationally expensive as it does not admit a closed-form solution.
Inspired by \cite{fang2015generalized}, we propose to add a proximal term  to improve the computational efficiency.  Let
$$
\begin{aligned}
\Psi^{t+1}  
 = & \argmin_{\Psi}  \text{vec}(\Psi)^\T\big\{(S\otimes I_p)^\T(S\otimes I_p)\big\}\text{vec}(\Psi)/2 - \text{vec}(Q)^\T\text{vec}(\Psi) + \text{vec}(\Lambda^t)^\T\text{vec}(\Psi) \\
  & + \frac{\rho}{2} \| \text{vec}(\Psi) - \text{vec}(\Phi^t)\|_2^2 + \frac{1}{2}\|\text{vec}(\Psi)  - \text{vec}(\Psi^t)  \|_G^2,
\end{aligned}
$$
where $\|\cdot\|_G$ denotes the matrix induced norm  $\|x\|_G = \sqrt{x^\T G x}$ for positive definite matrix $G$. Letting $G = \tau I_{p^2} - \big\{(S\otimes I_p)^\T(S\otimes I_p )\big\}$, where $\tau$ is greater than the largest eigenvalue of $\big\{(S\otimes I_p)^\T(S\otimes I_p) \big\}$, we have 
$$
\begin{aligned}
\Psi^{t+1}  = & \argmin_{\Psi} \frac{\rho+\tau}{2} \|\text{vec}(\Psi)\|_2^2 - \Big\{\text{vec}(Q) -\text{vec}(\Lambda^t) + (\rho+\tau)\text{vec}(\Phi^t) - (S\otimes I_p) \text{vec}(\Phi^t)\Big\}.
\end{aligned}
$$
Then, we have 
$$
\text{vec}(\Psi^{t+1})= \Big\{\text{vec}(Q) -\text{vec}(\Lambda^t) + (\rho+\tau)\text{vec}(\Phi^t) - (S\otimes I_p) \text{vec}(\Phi^t)\Big\}/(\rho+\tau).
$$
Thus, we have that by adding the proximal term, we have a closed-form solution for solving the $\Psi$-subproblem.


Next, considering the $\Phi$-subproblem, we have 
$$
\Phi^{t+1} =  \argmin_\Phi \cL_\rho(\Psi^{t+1},\Phi,\Lambda^{t}) = \argmin_\Phi \rho  \tr(\Phi^\T \Phi)/2 - \rho\tr(\Phi^\T \Psi^{t+1}) -\tr(\Phi^\T \Lambda^t) + \lambda\|\Phi\|_1.
$$
It is not difficult to see that this step admits a closed-form solution by soft thresholding: 
$$
\Phi^{t+1} = \text{shrinkage}(\Psi^{t+1} + \rho^{-1} \Lambda^t,\rho^{-1}\lambda),
$$
where $\text{shrinkage}(\cdot,\cdot):\mathbb{R}^{p\times p}\times \mathbb{R}\rightarrow \mathbb{R}^{p\times p}$ is the elementwise shrinkage operator that  for a matrix $A\in\RR^{p\times p}$ and $\tau>0$, the $(j,k)$-th entry $[\text{shrinkage}(A,\tau)]_{jk}= \text{sign}(A_{jk})\cdot\max\{0,|A_{jk}|-\tau\}$.

For the stopping criterion, we look at the primal and dual residuals. The primal residual is a measure of the feasibility for problem \eqref{eqn:adm}, which is defined as
$$
\eta^t_P = \big\|\Psi^t - \Phi^t\big\|_F.
$$
Meanwhile, the dual residual measures the convergence of the algorithm, where we take 
$$
\eta^t_D = \max\{\|\Psi^{t+1}-\Psi^t\|_F, \|\Phi^{t+1}-\Phi^t\|_F,\|\Lambda^{t+1}-\Lambda^t\|_F\}.
$$
In our implementation, we stop the algorithm when both primal and dual residuals are small that
$$
\max\{\eta^t_P,\eta_D^t\} \le 10^{-3}.
$$

We summarize the pseudo-code of the algorithm in Algorithm \ref{alg:1}. As we have derived the closed-form solutions for the  $\Psi$- and $\Phi$-subproblems in \eqref{eqn:admm}, which can be easily computed, Algorithm \ref{alg:1} can be implemented efficiently.
\begin{algorithm}
\caption{ADMM Algorithm to Estimate $\Psi$  \label{alg:1}}
\begin{algorithmic}[1]
\STATE \textbf{Input:} $S,Q\in\RR^{p\times p}$, $\lambda > 0$, $\Psi^0,\Phi^0,\Lambda^0\in\RR^{p\times p}$, $\rho,\tau>0$, $t=0$
\STATE \textbf{Output:}  $\widehat\Psi$
\WHILE {stopping criterion not satisfied}
\STATE  $\Psi^{t+1} \leftarrow \Big\{\text{vec}(Q) -\text{vec}(\Lambda^t) + (\rho+\tau)\text{vec}(\Phi^t) - (S\otimes I_p) \text{vec}(\Phi^t)\Big\}/(\rho+\tau)$
\STATE  $\Phi^{t+1} \leftarrow \text{shrinkage}(\Psi^{t+1} + \Lambda^t/\rho,\lambda/\rho)$
\STATE $\Lambda^{t+1} \leftarrow \Lambda^t + \rho(\Psi^{t+1} - \Phi^{t+1})$
\STATE $t\leftarrow t + 1$
\ENDWHILE
\STATE $\widehat\Psi \leftarrow\Psi^t$
\end{algorithmic}
\end{algorithm}

To conclude our method and algorithm, we make some remarks.  Our algorithm is designed for solve the matrix optimization problem  (\ref{eq:est}).  Our framework can be readily extended to a more general form of the weight loss function.  That is, by letting
$Z_A=A^{1/2}(\Sigma_X\Psi\Sigma_X-\Sigma_{YXX})A^{1/2}$
with some symmetric weighting matrix $A$, we define the loss
\begin{align*}
\mathcal{L}_A(\Psi)=\tr(Z_A^\T Z_A)=\tr\{ A(\Sigma_X\Psi\Sigma_X-\Sigma_{YXX})^\T A(\Sigma_X\Psi\Sigma_X-\Sigma_{YXX})\},
\end{align*} whose minimizer is also the principal Hessian matrix. 
Then $\mathcal{L}_A(\Psi)$ depends on $\Psi$ via
\begin{align*}
\tr\{(A\Sigma_X\Psi^\T \Sigma_{X})(A\Sigma_X\Psi \Sigma_X)\}-2 \tr(A\Sigma_X\Psi\Sigma_X A \Sigma_{YXX}).
\end{align*}
By choosing $A=\Sigma_X^{-1}$, the loss becomes (\ref{eq:est}).  Appropriate choice of $A$ may improve the speed and stability of the algorithm.  
For example,  
when   matrix $S$ is ill-conditioned, the choice of $A$ could be a viable device to solve the problem. 

\section{Numerical Studies}\label{s:sim} 
\subsection{Synthetic Data}
In this section, we conduct extensive numerical studies to demonstrate and validate the performance of our proposed method. We first conduct  investigations using synthetic data. In our simulation setup, we fix the sample size as $n = 100$, and we consider different dimensions for $p = 100$, 200 and 300. Meanwhile, we generate the design matrix $\XX = (x_1, x_2,...,x_n)^\T  \in\RR^{n\times p}$ by generating each sample $x_i\in\RR^p$ independently from a $p$-dimensional Gaussian distribution $X\sim N(0, \Sigma)$, where the covariance matrix $\Sigma$ is either the identity matrix, or a Toeplitz matrix, i.e. $\Sigma_{jk} = \rho^{|j-k|}$ for some $\rho\in(0,1)$. We then generate the noises $\epsilon_i$'s independently from a normal random variable $N(0,\sigma^2)$, and we consider different $\sigma$'s. 
To thoroughly compare the proposed method with other methods, we consider the following nine models:

\begin{enumerate}[align=left]
\item [Model 1:] $Y = X_1 + X_5 +\epsilon$, 
\item [Model 2:] $Y = 0.6X_1X_2 + 0.8X_4X_5+\epsilon$, 
\item [Model 3:] $Y = 0.6X_1X_2 + 0.8X_2X_3+\epsilon$, 
\item [Model 4:] $Y = 0.5X_1^2 + 0.9X_5X_8+\epsilon$, 
\item [Model 5:] $Y = X_1^2 + X_5X_8+ X_9^2 + \epsilon$, 
\item [Model 6:] $Y = X_1 + X_5+ X_1X_5 +\epsilon$, 
\item [Model 7:] $Y = 0.1X_1 + 0.1X_5+ X_1X_5 +\epsilon$, 
\item [Model 8:] $Y = X_1X_5+ X_2X_3\cdot\epsilon$,
\item [Model 9:] $Y = \sum_{j=1}^9X_jX_{j+1} + \epsilon$.
\end{enumerate}

In  Model 1, we consider the case where only the main effects are present.  This is a benchmarking case in the sense that there should be no false inclusion of the interactions for a valid detection method.  
In the next four 
 four models, we consider the cases where only interaction terms are presented under different scenarios.
With the interaction only models, we intend to show the advantage of the proposed one-step procedure. 
 Then, we consider two models where some hierarchical structures exist in Model  6 and Model 7 with different signal strength. Finally, we consider an example with the heteroscedasticity in Model 8, in which the conditional variance of the response variable is not homogeneous. Model 9 is a less sparse case.

We choose the tuning parameter by 10-fold cross-validation. We point out that after extensive numerical investigations, we find that our method is insensitive to the tuning parameter selection. We report the empirical true positive rate (TPR) and the false positive rate (FPR) by (\ref{set}) for each data generating scheme after repeating each scheme 200 times. In particular, let $\cI$ be the true set of interaction terms, and let $\widehat\cI$ be the interaction terms selected by the estimator $\widehat\Psi$. TPR and FPR are defined as
$$
\text{TPR} = \frac{|\cI\cap\widehat{\cI}|}{|\cI|}, \text{ and }\ \text{FPR} = \frac{|\widehat{\cI}\backslash \cI|}{{d\choose 2}+d -|\cI|}
$$
respectively. 

We compare our method with the interaction pursuit (IP) method \citep{Fan2016}, which is two-step method with a screening procedure as the first step, and the RAMP method \citep{hao2018model}, which is based on a hierarchical structure. We report the results in Tables \ref{tab:1}-\ref{tab:5}. 

Firstly, as seen in Table \ref{tab:1} for Model 1, our one-step method performs very well with very little false inclusion when no interactions are present.  Other competing methods also are performing very well in the sense of little of no false inclusions. 
In all other models for detecting the contributing interactions, 
 we see that the proposed method outperforms the other two methods by large margins in all settings except Model 6, where a hierarchical structure is presented with good signal strength, which favors the RAMP method when the main effect is correctly detected first. Nevertheless, we see that in Model 7, though a hierarchical model is presented, due to the low signal strength of the main effect of $X_1$ and $X_5$, our method still outperforms the RAMP method substantially. This clearly demonstrates the advantage of our one-step estimation based interaction detection method. 

In all tables, we report the averaged running time in seconds, where all experiments are conducted on an iMac with 3.2 GHz Intel Core i5 Processor and 16 GB memory. It is seen that the IP method is most computationally efficient, which is expected as it is a screening based method. Meanwhile, from the computational perspective, our method's efficiency is similar to the RAMP method. However, we point out that our method directly deals with the principal Hessian matrix and do not impose any hierarchical structure. In comparison, the RAMP method imposes a hierarchical structure, and essentially deals with vectors.

\subsection{Real Data}
We further apply the proposed method to analyze the GPL96 microarray dataset analyzed in \cite{mccall2010frozen,wu2013chip,fang2017mining}. This dataset contains $13,182$ samples of more than 2,000 biological contexts generated from Affymetrix Human 133A (GPL96) arrays, and each sample has $12,704$ genes. The data set was preprocessed and normalized using frozen-RMA \citep{mccall2010frozen} to reduce batch effects. We use $n_1 = 235$ samples of breast tumor and $n_2 = 537$ normal samples in the analysis. To improve the efficiency in our demonstration, we conduct some pre-screening. In particular,  we use a plug-in estimator $\tilde{\Psi} = S^-QS^-$ to estimate $\Psi$, where $S^-$ denotes the Moore-Penrose pseudo-inverse of $S$. We then screen out the variables where the $\ell_1$-norm of the corresponding columns of  $\tilde{\Psi}$ are small. In our analysis, we  screen the genes and keep $p = 1,000$ genes. 

We treat the disease status as responses, and randomly split the dataset into a training set and a testing set. Each training set contains 100 samples from the breast tumor group and 150 samples from the normal group. We repeat the random split 100 times. We first report the averaged testing errors and their standard errors  in Table~\ref{tab:r1}. It is seen that the proposed method achieves better testing errors than the other two methods. We further provide the  five most selected interaction terms in Table~\ref{tab:r2}. It is seen that BRCA1 and BRCA2 are frequently selected by our proposed method including the interaction between these two genes, while they are less frequently selected by the other two methods. It is well known in literature that BRCA1 and BRCA2 are of fundamental importance for breast tumor as shown in \cite{king2003breast}, and this demonstrates the potential usefulness of the proposed method in analyzing real data.

Furthermore, we consider another disease Ewing Sarcoma, where we have $n_1 = 57$ samples. Following same procedure, each time, we randomly select 33 (57.9\%) of these samples and 150 samples from the normal group as training samples and the rest as testing samples. We repeat the random splitting 100 times and report the average classification error rates in Table \ref{tab:re}. It is seen that the RAMP method does not perform well in this case, and our method performs better than the IP method. This demonstrates the importance of including interaction terms in the model.


\renewcommand{\baselinestretch}{\vm}
\normalsize

\section{Some Theoretical Analysis}\label{s5}
\subsection{Non-asymptotic Results}
We now perform some theoretical analysis of the estimator (\ref{eq:est}).  Outline of the proofs is provided in the Appendix of the paper. 
Denote by $\Psi^*$ the unknown truth of the principal Hessian matrix, and ${\cal U}=\supp(\Psi^*)$ the support of $\Psi$, and $q=|{\cal U}|$ be the cardinality of ${\cal U}$.
Denote by $\|A\|_\infty$ the max-norm of the matrix $A$.

The following lemma establishes a non-asymptotic result on the support  of $\widehat \Psi$.

\begin{lemma}\label{la:1}
Let ${\cal C}(\theta^*)=\{\Psi:  \|\Psi_{{\cal U}^c}\|_1\leq 3  \|\Psi_{\cal U}-\Psi_{\cal U}^*\|_1\}$ be the cone depending on the truth $\Psi^*$. We have that $\widehat\Psi \in {\cal C}(\theta^*)$ provided that the tuning parameter $\lambda$ satisfies $2\|S \Psi^* S-Q\|_\infty \leq\lambda/2$.
\end{lemma}

We also have the following lemma containing a non-asymptotic oracle's inequality.

\begin{lemma}\label{la:2}
Let $\widehat\Delta =\widehat\Psi-\Psi^*. $
 If $2\|S \Psi^* S-Q\|_\infty <\lambda/2$, then
\begin{align*}
2\tr(\widehat\Delta S \widehat \Delta S)+\lambda\|\widehat\Psi_{\cal U}-\Psi_{\cal U}^*\|_1+\lambda \|\widehat \Psi_{{\cal U}^c}\|_1\leq 4 \lambda\|\widehat\Psi_{\cal U}-\Psi_{\cal U}^*\|.
\end{align*}
\end{lemma}

To establish an estimation error bound of $\widehat\Psi$, we need the following condition.

\begin{condition}\label{con:1}
(Restricted eigenvalue condition)
Let $\Delta=\Psi-\Psi^*$. For all $\Psi\in{\cal C}(\Psi^*)$,   \begin{align}\label{eq:rec}
\vecc^\T (\Delta)(S\otimes S) \vecc(\Delta)\ge \gamma \|\vecc(\Delta)\|_2^2.
\end{align}
\end{condition}

Then we have the following theorem, establishing a key non-asymptotic result on  the estimating error bound of $\widehat\Psi$.
\begin{theorem}\label{tm:1}
Suppose that  the restricted eigenvalue Condition \ref{con:1} holds. If  $2\|S \Psi^*S -Q\|_\infty \leq\lambda/2$,  then
$
\|\widehat \Psi-\Psi^*\|_1 \leq  4 \lambda q /\gamma^2$.
\end{theorem}

The main implication from Theorem \ref{tm:1} is the good performance non-asymptotically in the sense of small error bound. 
The restricted eigenvalue Condition \ref{con:1} essentially requires that on the set ${\cal C}$, the smallest eigenvalue of $S\otimes S$ is strictly bounded away from 0. We remark that the condition (\ref{eq:rec}) here is essentially not stronger than the analogous one with high-dimensional linear model $Y=X\beta+\epsilon$ for sparse estimator of~$\beta$. For example, as in \cite{Hastieetal_2015} and \cite{Buhlmannvan_2011}, the Lasso estimator requires 
$
(\beta-\beta^*)^\T  S ( \beta-\beta^*)\ge \gamma_1 \| \beta-\beta^*\|_2^2$ 
for all $\beta$ in an analogous restricted set depending on the unknown truth $\beta^*$.  Then by observing the fact that the  eigenvalues of $ H_1=S\otimes S$ are the products of the  eigenvalues of $S$,  we see that Condition \ref{con:1} is essentially of the same kind as those on the sample covariance matrix.   This can be viewed as a merit of our method. 
In contrast, if one considers an alternative approach incorporating all interactions in a penalized regression,  then the corresponding restricted eigenvalue condition is required on the matrix $H_2=n^{-1}\sum_{i=1}^n  w_iw_i^\T$ with $w_i\in{\mathbb R}^{p(p+1)/2}$ containing all quadratic terms of the $i$th observation.  Clearly, a much larger sample covariance matrix $H_2$ is directly involved with fourth order moments; and conditions on its restricted eigenvalue  are seen as more stringent.  
Additionally, comparing with hierarchical approaches using two-step procedures,  the main difference is that our condition does not require condition on the model matrix to ensure correctly identifying the linear effects; see, for example, the irrepresentable condition of \cite{hao2018model}. 

\subsection{Asymptotic Results} \label{theory}

The restricted eigenvalue condition (\ref{eq:rec}) is known important for establishing the error bounds of penalized estimators in regression problems; see, among others,   \cite{Negahban2012} and  \cite{Hastieetal_2015}.  Since our estimator (\ref{eq:est})  is the minimizer of the sum a quadratic loss and $\ell_1$ penalty, this type of the restricted eigenvalue condition is expected.
It is known in the literature that in an asymptotic setting assuming that the entries of $X_i$ following some distributions satisfying some condition on the tail probabilistic behavior, then the condition (\ref{eq:rec}) holds with probability tending to 1. We refer to the discussion in   \cite{Negahban2012} and results in \cite{Raskutti2010}, \cite{Raskutti2011}, and \cite{Rudelson2011}.

As in Lemma \ref{la:1} and Theorem \ref{tm:1}, the condition  $2\|S \Psi^* S-Q\|_\infty \leq \lambda/2$ imposes the requirement on the tuning parameter $\lambda$.
By their definitions, $Q$ is an estimator of $\Sigma_{YXX}$, and $S$ is an estimator of $\Sigma_X$. As in an asymptotic setting with $n\to\infty$ and $p\to\infty$, they all converge to the truths element-wise in probability under appropriate conditions on the tail distribution of $Y$ and $X$.  Thus, the tuning parameter is allowed going to $0$ as $n\to\infty$, ensuring that $\widehat{\Psi} \in {\cal C}(\Psi^*)$, and $\|\widehat\Delta\|_1\to 0$ in probability, i.e. consistency of~$\widehat \Psi$.

Formally, for showing the asymptotic properties of $\widehat\Psi$, we impose the following condition.
\begin{condition}\label{con:2}
The random vectors $\{(Y_i, X_i^\T)^\T\}_{i=1}^n$ are independent and identically distributed.  The largest eigenvalue of the $X_i$ is strictly bounded away from infinity.  Both $Y_i$ and $X_{ij}$ $(j=1,\dots,p)$ satisfy exponential tail property, i.e., there exist constants $b_1$, $b_2$, $c_1$ and $c_2$ such that
\[
\PP(|Y_i|>t)\leq \exp\{-(t/b_1)^{c_1}\}\quad\text{ and }\quad \PP(|X_{ij}|>t)\leq \exp\{-(t/b_2)^{c_2}\}.
\]
\end{condition}

Condition \ref{con:2} ensures the concentration property of $\|Q-\Sigma_{YXX}\|_\infty$. So that by choosing $\lambda=C \sqrt{\frac{\log p}{n}}$,  $2\|Q-\Sigma_{YXX}\|_\infty\leq \lambda/2$ holds with probability at least $1-c_3\exp(-c_4 n^{c_5}\lambda^2)$ when $\log(p)=n^{\eta}$ with some constants $C, c_3,c_4,c_5, \eta$. 

Then we have the following asymptotic result on the error bound of $\widehat \Psi$.

\begin{theorem}\label{tm:2}
Suppose that $\Psi^*$ is sparse with support ${\cal U}$, and the cardinality $|{\cal U}|=q$.   Suppose additionally that $\|\Psi^*\|_\infty=O\left(\sqrt{{n}/{\log p}}\right)$, $\|\Psi^*\Sigma_X\|_\infty=O(1)$.  Then under Condition \ref{con:2}, by choosing  $\lambda=C \sqrt{\frac{\log p}{n}}$,  $\|\widehat \Psi-\Psi^*\|_1=\mathcal{O}_p(q \sqrt{\log p/n})$.
\end{theorem}

To ensure that zero components of $\Psi^*$ are correctly estimated as zero by $\widehat \Psi$, we need the following irrepresentable condition.

\begin{condition}\label{con:3}
(Irrepresentable condition) Let $\Lambda=S\otimes S=\begin{pmatrix}\Lambda_{11}&\Lambda_{12}\\
\Lambda_{21}&\Lambda_{22}\end{pmatrix}$ with $\Lambda_{11}$ and $\Lambda_{22}$ corresponding to the partition of $\vecc(\Psi^*_{\cal U})$ and $\vecc(\Psi^*_{{\cal U}^c})$. Then $\max_j\|\Lambda_{11}^{-1}\Lambda_{12,j}\|_1\leq 1-\alpha$ for some $\alpha>0$, where $\Lambda_{12,j}$ is the $j$-th column of $\Lambda_{12}$.

\end{condition}

\begin{theorem}\label{tm:3}
By choosing  $\lambda=C \sqrt{\frac{\log p}{n}}$ and assuming Conditions \ref{con:2} and \ref{con:3}, $\widehat\Psi_{{\cal U}^c}=0$ with probability tending to 1. Furthermore, if $|\min(\Psi^*_{\cal U})|>\|\lambda\Lambda_{11}^{-1}\|_\infty$, then  $\widehat\Psi_{{\cal U}}$ is bounded away from zero with probability tending to 1.
\end{theorem}

With Theorem \ref{tm:2} showing the consistence of $\widehat\Psi$, and Theorem \ref{tm:3} showing the correctness in identifying the nonzero component in $\Psi$, we demonstrate the validity of the our method in detecting the interactions.

\renewcommand{\baselinestretch}{\vt}
\normalsize

 \begin{table}
\centering
\caption{\small Quantitative performance of the interaction detection methods for Model 1 and Model 2. We report the averaged true positive and false positive discovery rates after repeating each simulation setup 200 times with $n=100$. Note that $\rho = 0$ refers to the case where $\bSigma$ is the identity matrix.}

\begin{adjustbox}{width=\textwidth}
\begin{tabular}{r*{ 12}{r}}
\toprule
& & &  \multicolumn{3}{c}{$d$ = 100} & \multicolumn{3}{c}{$d$ = 200}  & \multicolumn{3}{c}{$d$ = 300} \\
 \cmidrule{4-12}
Model  & ($\rho$, $\sigma$) &  Method & TPR &  FPR & Time(s) & TPR &  FPR & Time(s) & TPR &  FPR & Time(s)  \\
 \midrule
   1 & (0,\ 0.1) & ADMM  & NA & 0.03\% & 0.41(0.05) & NA & 0.01\% & 1.27(0.13) & NA & 0.00\% & 2.13(0.25)\\
& & IP & NA & 0.15\% & 0.12(0.02) & NA & 0.11\% & 0.20(0.03) & NA & 0.03\% & 0.34(0.04)\\
& & RAMP  & NA & 0.00\% & 0.39(0.04)  & NA & 0.00\% & 1.15(0.10) & NA & 0.00\% & 1.95(0.19)\\
 & (0,\ 2.0) & ADMM  & NA & 0.04\% & 0.38(0.04) & NA & 0.01\% & 1.15(0.07) & NA & 0.00\% & 1.99(0.14)\\
& & IP  & NA & 0.14\% &  0.09(0.01) & NA & 0.02\% & 0.17(0.03) & NA & 0.00\% & 0.29(0.07)\\
& & RAMP   & NA & 0.00\% & 0.37(0.06) &  NA & 0.00\% & 1.20(0.13) & NA & 0.00\% & 2.04(0.22)\\
& (0.2,\ 1.0) & ADMM  & NA & 0.05\% & 0.44(0.03) & NA & 0.03\% & 1.33(0.17) & NA & 0.01\% & 2.25(0.29)\\
& & IP & NA & 0.08\% & 0.14(0.01) & NA & 0.02\% & 0.23(0.03) & NA & 0.00\% & 0.36(0.03)\\
& & RAMP  & NA & 0.00\% & 0.35(0.05) & NA & 0.00\% & 1.05(0.13) & NA & 0.00\% & 1.92(0.31)\\
 & (0.2,\ 2.0) & ADMM  & NA & 0.05\% & 0.46(0.03) & NA & 0.02\% & 1.23(0.18) & NA & 0.00\% & 2.07(0.30)\\
& & IP & NA & 0.12\% & 0.10(0.01)  & NA & 0.03\% & 0.18(0.02) & NA & 0.00\% & 0.31(0.02)\\
& & RAMP  & NA & 0.00\% & 0.50(0.10) & NA & 0.00\% & 1.15(0.20) & NA & 0.00\% & 2.23(0.22)\\
\midrule
  2 & (0,\ 0.1) & ADMM  & 99.0\% & 0.08\% & 0.44(0.08) & 96.0\% & 0.09\% & 0.95(0.12) & 92.5\% & 0.13\%  & 2.31(0.32)\\
& & IP & 82.0\% & 0.06\%  & 0.08(0.01)    &  67.0\% & 0.08\% & 0.23(0.04) & 66.0\% & 0.02\% & 0.34(0.04)  \\
& & RAMP  & 1.00\% & 0.03\% & 0.48(0.10) & 0.00\% & 0.06\% & 0.99(0.13) & 0.00\% & 0.01\%  & 2.03(0.24) \\
& (0,\ 1.0) & ADMM  & 98.5\% & 0.12\% & 0.40(0.05) & 92.0\% & 0.13\% & 1.22(0.10) & 90.0\% & 0.17\% & 2.18(0.29)\\
& & IP & 54.0\% & 0.01\%    & 0.07(0.02)   &  47.5\% & 0.04\% & 0.18(0.03) & 42.5\% & 0.04\% & 0.41(0.05) \\
& & RAMP  & 2.00\% & 0.22\% & 0.38(0.02)   & 0.00\% & 0.06\% & 1.15(0.09) & 0.00\% & 0.00\% & 2.30(0.25) \\
& (0.1,\ 0.1) & ADMM  & 96.5\% & 0.14\%  & 0.35(0.09) & 93.5\% & 0.18\% & 0.93(0.14) & 91.0\% & 0.11\% & 2.20(0.28)\\
& & IP & 73.0\% & 0.19\% & 0.09(0.02) &  68.5\% & 0.11\%  & 0.20(0.03) & 65.0\% & 0.08\% & 0.38(0.05)  \\
& & RAMP  & 0.00\% & 0.00\% & 0.29(0.11) & 0.00\% & 0.00\% & 1.23(0.12) & 0.00\% & 0.00\% & 2.37(0.30) \\
& (0.1,\ 1.0) & ADMM  & 95.0\% & 0.30\% & 0.37(0.03) & 91.5\% & 0.26\% & 1.20(0.18) & 89.0\% & 0.13\% & 2.34(0.31)\\
& & IP & 49.5\% & 0.18\% & 0.10(0.03)&  40.5\%  & 0.10\% & 0.25(0.03) & 28.0\% & 0.06\% & 0.31(0.03)  \\
& & RAMP  & 0.00\% & 0.00\% & 0.45(0.02) & 0.00\% & 0.00\% & 1.27(0.23) & 0.00\% & 0.00\% & 2.48(0.43) \\
  \bottomrule
 \end{tabular}
 \end{adjustbox}
 \label{tab:1}
 \end{table}

  \begin{table}
\centering
\caption{Quantitative performance of the interaction detection methods for Model 3 and Model 4. We report the averaged true positive and false positive discovery rates after repeating each simulation setup 200 times with $n=100$. Note that $\rho = 0$ refers to the case where $\bSigma$ is the identity matrix.}

\vspace{0.5cm}

\begin{adjustbox}{width=\textwidth}
\begin{tabular}{r*{ 12}{r}}
\toprule
& & &  \multicolumn{3}{c}{$d$ = 100} & \multicolumn{3}{c}{$d$ = 200}  & \multicolumn{3}{c}{$d$ = 300} \\
 \cmidrule{4-12}
Model  & ($\rho$, $\sigma$) &  Method & TPR &  FPR & Time(s) & TPR &  FPR & Time(s) & TPR &  FPR & Time(s)  \\
 \midrule
   3 & (0.1,\ 0.1) & ADMM  & 99.5\% & 0.10\% & 0.36(0.03) & 95.0\% & 0.08\% & 1.03(0.05) & 94.5\% & 0.13\% & 2.04(0.10) \\
& & IP & 93.5\% & 0.19\%  & 0.08(0.01)   &  82.5\% & 0.05\% & 0.17(0.02)  &  80.5\% & 0.03\% & 0.28(0.02)  \\
& & RAMP  & 2.00\% & 0.02\% & 0.38(0.02) & 2.00\% & 0.01\% & 1.09(0.06) & 2.00\% & 0.00\% & 2.23(0.15) \\
& (0.1,\ 1.0) & ADMM  & 96.5\% & 0.29\% & 0.46(0.05)  & 93.0\% & 0.13\% & 0.96(0.17) & 91.5\% & 0.08\% & 2.10(0.14)\\
& & IP & 76.0\% & 0.09\% & 0.15(0.02)   &  68.5\% & 0.03\% & 0.25(0.05)  & 62.0\% & 0.02\% & 0.37(0.04)  \\
& & RAMP  & 0.00\% & 0.00\%  & 0.43(0.06)   & 1.00\% & 0.00\%  & 0.99(0.10) & 0.00\% & 0.00\% & 2.39(0.18) \\
& (0.4,\ 0.1) & ADMM  & 100.0\% & 0.20\% & 0.40(0.03) & 99.0\% & 0.15\%  & 1.35(0.12)  & 96.0\% & 0.07\% & 2.36(0.28)\\
& & IP & 97.0\% & 0.10\% & 0.07(0.01) &   91.5\% & 0.08\% & 0.15(0.04)  & 89.0\% & 0.05\% & 0.24(0.06)  \\
& & RAMP  & 2.00\% & 0.00\% & 0.49(0.05)  & 0.00\% & 0.00\% & 1.34(0.14) & 3.00\% & 0.00\% & 2.59(0.30) \\
& (0.4,\ 1.0) & ADMM  & 99.0\% & 0.40\% & 0.43(0.07)  & 99.0\% & 0.19\% & 1.31(0.16)& 97.0\% & 0.10\% & 2.39(0.21)\\
& & IP & 92.0\% & 0.10\% & 0.06(0.02) &  89.5\% & 0.08\% & 0.16(0.03)  & 86.0\% & 0.02\% & 0.25(0.04)  \\
& & RAMP  & 2.00\% & 0.00\% & 0.37(0.13)   & 0.50\% & 0.00\% & 0.98(0.10) & 0.50\% & 0.00\% & 2.27(0.18) \\
\midrule
  4 & (0,\ 1.0) & ADMM  & 98.5\% & 0.12\% & 0.24(0.05) & 96.0\% & 0.09\% & 1.17(0.11)  & 95.5\% & 0.03\% & 2.19(0.15)\\
& & IP & 89.0\% & 0.08\% & 0.07(0.02)  &  82.5\% & 0.04\% & 0.20(0.03) & 73.0\% & 0.02\% & 0.38(0.03)  \\
& & RAMP  & 3.50\% & 0.00\% & 0.25(0.03)  & 1.00\% & 0.06\% & 1.29(0.24) & 0.00\% & 0.01\% & 2.10(0.37) \\
& (0,\ 1.5) & ADMM  & 98.0\% & 0.13\% & 0.33(0.06)  & 94.0\% & 0.09\% & 1.21(0.17) & 91.5\% & 0.04\%  & 2.15(0.20)  \\
& & IP & 74.0\% & 0.09\%   & 0.06(0.04)   &  62.0\% & 0.05\% & 0.18(0.03)  & 52.0\% & 0.03\% & 0.45(0.04)  \\
& & RAMP  & 0.50\% & 0.00\% & 0.27(0.05)  & 2.00\% & 0.00\% & 1.36(0.15) & 0.00\% & 0.00\% & 2.32(0.26)  \\
& (0.05,\ 0.5) & ADMM  & 98.0\% & 0.15\% & 0.31(0.07)  & 95.0\% & 0.09\%  & 1.08(0.15)   &   93.0\% & 0.13\% & 2.51(0.24)\\
& & IP & 94.5\% & 0.12\% & 0.06(0.01) &  91.5\% & 0.10\% & 0.15(0.02)  & 89.0\% & 0.06\% & 0.33(0.03) \\
& & RAMP  & 9.00\% & 0.00\% & 0.28(0.03) & 6.00\% & 0.00\% & 1.23(0.12) & 0.00\% & 0.00\% & 2.40(0.16) \\
& (0.15,\ 1.0) & ADMM  & 98.5\% & 0.18\% & 0.24(0.04) & 94.0\% & 0.13\% & 1.19(0.10) & 93.0\% & 0.06\% & 2.20(0.27)\\
& & IP & 90.5\% & 0.13\% & 0.08(0.02) &  81.5\% & 0.06\% & 0.16(0.03) & 76.5\% & 0.02\%  & 0.29(0.03)\\
& & RAMP  & 8.00\% & 0.00\% &  0.19(0.03) & 3.00\%  & 0.00\% & 1.05(0.24) & 2.50\%  & 0.00\% & 2.35(0.23)\\
  \bottomrule
 \end{tabular}
  \end{adjustbox}
 \label{tab:2}
 \end{table}

   \begin{table}
\centering
\caption{Quantitative performance of the interaction detection methods for Model 5 and Model 6. We report the averaged true positive and false positive discovery rates after repeating each simulation setup 200 times with $n=100$. Note that $\rho = 0$ refers to the case where $\bSigma$ is the identity matrix.}

\begin{adjustbox}{width=\textwidth}
\begin{tabular}{r*{ 12}{r}}
\toprule
& & &  \multicolumn{3}{c}{$d$ = 100} & \multicolumn{3}{c}{$d$ = 200}  & \multicolumn{3}{c}{$d$ = 300} \\
 \cmidrule{4-12}
Model  & ($\rho$, $\sigma$) &  Method & TPR &  FPR & Time(s) & TPR &  FPR & Time(s) & TPR &  FPR & Time(s)  \\
 \midrule
  5 & (0,\ 0.1) & ADMM  & 94.7\% & 0.17\% &  0.33(0.02)  & 92.0\% & 0.09\% &  1.14(0.04)  & 90.3\% & 0.05\%  &  2.27(0.13)\\
& & IP & 80.7\% & 0.14\%  &  0.05(0.01)  &  91.5\% & 0.10\% &  0.18(0.02) & 70.3\% & 0.04\% &  0.30(0.04)  \\
& & RAMP  & 9.33\% & 0.00\% &  0.35(0.03)  & 9.67\% & 0.00\% & 1.19(0.10) & 10.7\% & 0.00\% & 2.15(0.18) \\
 & (0,\ 1.0) & ADMM  & 92.0\% & 0.12\% & 0.29(0.04) & 91.3\% & 0.07\% & 1.24(0.14) & 89.7\% & 0.05\% & 2.38(0.21)\\
& & IP & 78.3\% & 0.11\%  & 0.08(0.01)  & 75.0\% & 0.06\% & 0.15(0.02) & 68.7\% & 0.03\% & 0.26(0.03)  \\
& & RAMP  & 8.00\% & 0.00\% & 0.24(0.02) & 3.00\% & 0.00\% & 1.32(0.13) & 6.67\% & 0.00\% & 2.41(0.24) \\
 & (0.2,\ 0.1) & ADMM  & 95.3\% & 0.31\% & 0.42(0.02)  & 94.0\% & 0.10\% & 1.20(0.17) &  93.7\% & 0.04\% & 2.48(0.26) \\
& & IP & 81.3\% & 0.28\% & 0.11(0.01) &  72.3\% & 0.09\% & 0.16(0.02) & 71.0\% & 0.03\% & 0.34(0.02)  \\
& & RAMP  & 16.0\% & 0.00\% & 0.39(0.05)   & 11.3\% & 0.00\% & 1.37(0.16) & 0.00\% & 0.01\% 
& 2.66(0.47)
\\
& (0.2,\ 1.0) & ADMM  & 93.7\% & 0.11\% & 0.36(0.04) & 92.7\% & 0.08\%  & 1.24(0.15)& 91.5\% & 0.04\% & 2.53(0.31)\\
& & IP & 79.7\% & 0.08\% & 0.07(0.02) &  77.3\% & 0.06\%   & 0.19(0.02)  & 69.7\% & 0.04\%  & 0.33(0.03)\\
& & RAMP  & 10.7\% & 0.00\% & 0.31(0.04) &  8.33\% & 0.00\% & 1.40(0.18) & 7.33\% & 0.00\% & 2.45(0.39) \\
 \midrule
  6 & (0,\ 1.0) & ADMM  & 98.5\% & 0.23\% & 0.24(0.03) & 93.0\% & 0.19\% & 1.18(0.13)  & 91.5\% & 0.10\%  & 2.34(0.32)  \\
& & IP & 98.5\% & 0.20\% & 0.08(0.02) &  92.5\% & 0.12\% & 0.21(0.03) & 92.0\% & 0.08\% & 0.31(0.04)  \\
& & RAMP  & 99.5\% & 0.00\% & 0.32(0.03)  & 100.0\% & 0.06\% & 1.50(0.22) & 100.0\% & 0.00\% & 2.24(0.19)\\
& (0,\ 1.5) & ADMM  & 92.0\% & 0.31\%  & 0.43(0.08)  & 88.5\% & 0.16\%  & 1.52(0.18)  & 82.0\% & 0.04\% & 2.40(0.24)    \\
& & IP & 91.5\% & 0.15\% & 0.06(0.02)   &  86.0\% & 0.11\% & 0.17(0.03)   & 80.0\% & 0.03\% & 0.34(0.05)    \\
& & RAMP  & 96.5\% & 0.00\% & 0.29(0.04)   & 95.5\% & 0.00\% & 1.34(0.27)   & 94.0\% & 0.00\% & 2.39(0.25)   \\
& (0.2,\ 1.0) & ADMM  & 96.0\% & 0.23\% & 0.36(0.05)   & 95.0\% & 0.14\% & 1.49(0.20)   & 92.5\% & 0.08\% & 2.61(0.38)  \\
& & IP & 97.0\% & 0.12\% & 0.09(0.02) &  94.5\%     & 0.08\% & 0.18(0.03)  & 91.0\% & 0.05\% & 0.29(0.03)  \\
& & RAMP  & 100.0\% & 0.00\% & 0.41(0.06)  & 100.0\% & 0.00\% & 1.33(0.24) & 99.0\% & 0.00\% & 2.45(0.36) \\
& (0.4,\ 1.0) & ADMM  & 97.5\% & 0.19\% & 0.42(0.06) & 95.5\% & 0.10\% & 1.38(0.22) & 95.0\% & 0.04\% & 2.65(0.31)\\
& & IP & 98.0\% & 0.11\% & 0.07(0.02)  &  96.0\% & 0.06\% & 0.20(0.03) & 93.5\% & 0.01\% & 0.32(0.04)  \\
& & RAMP  & 100.0\% & 0.00\% & 0.35(0.05) & 100.0\% & 0.00\% & 1.39(0.30) & 100.0\% & 0.00\% & 2.50(0.39) \\
  \bottomrule
 \end{tabular}
   \end{adjustbox}
 \label{tab:3}
 \end{table}

    \begin{table}
\centering
\caption{Quantitative performance of the interaction detection methods for Model 7 and Model 8. We report the averaged true positive and false positive discovery rates after repeating each simulation setup 200 times with $n=100$. Note that $\rho = 0$ refers to the case where $\bSigma$ is the identity matrix.}

\begin{adjustbox}{width=\textwidth}
\begin{tabular}{r*{ 12}{r}}
\toprule
& & &  \multicolumn{3}{c}{$d$ = 100} & \multicolumn{3}{c}{$d$ = 200}  & \multicolumn{3}{c}{$d$ = 300} \\
 \cmidrule{4-12}
Model  & ($\rho$, $\sigma$) &  Method & TPR &  FPR & Time(s) & TPR &  FPR & Time(s) & TPR &  FPR & Time(s)  \\
 \midrule
7 & (0,\ 1.0) & ADMM  & 100.0\% & 0.29\%  & 0.31(0.04)   & 100.0\% & 0.19\% & 1.22(0.12) & 97.5\% & 0.08\% & 2.38(0.35)\\
& & IP & 98.5\% & 0.27\% & 0.08(0.02) & 95.5\% & 0.13\% & 0.17(0.02) & 90.0\% & 0.04\% & 0.36(0.04)\\
& & RAMP  & 1.00\% & 0.00\%  & 0.36(0.06)   & 2.50\% & 0.00\% & 1.31(0.11)  & 2.00\% & 0.00\%
& 2.50(0.42) \\
 & (0,\ 1.5) & ADMM  & 97.0\% & 0.34\%  & 0.41(0.06)  &  95.5\% & 0.17\%  & 1.30(0.21)
 & 94.5\% & 0.07\% & 2.43(0.28)  \\
& & IP  & 94.0\% & 0.29\% & 0.08(0.02) & 92.0\% & 0.16\% & 0.21(0.03) & 89.5\% & 0.06\% & 0.37(0.06)\\
& & RAMP   & 1.50\% & 0.00\% & 0.37(0.04)    & 2.00\% & 0.00\%  & 1.43(0.32) & 1.00\% & 0.00\%  & 2.62(0.38)\\
& (0.2,\ 1.0) & ADMM  & 100.0\% & 0.23\% & 0.39(0.03) & 99.5\% & 0.13\%  & 1.40(0.18) & 97.0\% & 0.06\%   & 2.44(0.31)\\
& & IP & 97.0\% & 0.24\%   & 0.07(0.02)  & 94.0\%  & 0.12\% & 0.19(0.03) & 90.5\% & 0.05\%  & 0.32(0.04)\\
& & RAMP  & 1.50\% & 0.00\% & 0.40(0.03)   & 2.50\% & 0.00\%  & 1.35(0.25) & 2.00\% & 0.00\% & 2.23(0.40)\\
 & (0.2,\ 1.5) & ADMM  & 96.5\% & 0.25\%   & 0.33(0.04)   & 95.0\% & 0.17\%  & 1.24(0.18) & 94.0\% & 0.06\%  & 2.48(0.29)\\
& & IP & 93.0\% & 0.27\%  & 0.09(0.02) & 91.5\% & 0.13\%  & 0.16(0.03)& 88.0\% & 0.04\% & 0.31(0.04)\\
& & RAMP  & 1.00\% & 0.00\% & 0.38(0.06) & 1.50\% & 0.00\% & 1.32(0.24)  & 1.00\% & 0.00\%  & 2.35(0.33)\\
\midrule
8  & (0,\ 1.0) & ADMM  & 97.0\% & 0.31\% & 0.27(0.03) & 96.5\% & 0.18\% & 1.40(0.18) & 94.5\% & 0.05\% & 2.51(0.32)\\
& & IP & 75.5\%  & 0.29\% & 0.06(0.02)  & 60.0\% & 0.17\% & 0.18(0.03)  & 57.5\% & 0.08\% & 0.39(0.05)\\
& & RAMP  & 0.00\% & 0.00\% & 0.25(0.04) & 1.00\% & 0.00\%  & 1.37(0.29) & 0.00\% & 0.00\% & 2.40(0.34)\\
& (0,\ 1.25) & ADMM  & 95.0\% & 0.28\% & 0.40(0.06) & 93.5\% & 0.16\% & 1.28(0.26) & 91.0\% & 0.08\% & 2.48(0.41)\\
& & IP  & 48.5\% & 0.23\%  & 0.07(0.02) & 39.0\% & 0.13\%  & 0.17(0.02) & 35.0\% & 0.06\%   & 0.37(0.06)\\
& & RAMP   & 1.00\% & 0.00\% & 0.36(0.05) & 2.50\% & 0.00\%   & 1.29(0.17)  & 1.00\% & 0.00\% & 2.36(0.38)\\
& (0.2,\ 1.0) & ADMM  & 95.5\% & 0.35\% & 0.32(0.04)  & 92.5\% & 0.23\% & 1.19(0.15)   & 91.5\% & 0.12\% & 2.58(0.31)\\
& & IP & 67.5\% & 0.28\% & 0.08(0.02)   & 54.0\% & 0.20\% & 0.15(0.03)   & 52.5\% & 0.13\% & 0.33(0.05)\\
& & RAMP  & 1.00\% & 0.00\% & 0.36(0.06) & 1.00\% & 0.00\% & 1.41(0.16)  & 1.50\% & 0.00\% & 2.43(0.41)\\
 & (0.2,\ 1.25) & ADMM  & 94.0\% & 0.29\% & 0.42(0.06) & 92.0\% & 0.17\% & 1.38(0.19) & 90.5\% & 0.09\% & 2.43(0.44)\\
& & IP & 64.0\% & 0.25\% & 0.09(0.02)  & 52.0\% & 0.03\% & 0.20(0.04) & 50.5\% & 0.08\% & 0.33(0.05)\\
& & RAMP  & 1.00\% & 0.00\% & 0.30(0.04) & 2.50\% & 0.00\% & 1.24(0.26) & 2.00\% & 0.00\% & 2.39(0.35)\\
  \bottomrule
 \end{tabular}
   \end{adjustbox}
 \label{tab:4}
 \end{table}

    \begin{table}
\centering
\caption{Quantitative performance of the interaction detection methods for Model 9. We report the averaged true positive and false positive discovery rates after repeating each simulation setup 200 times with $n=100$. Note that $\rho = 0$ refers to the case where $\bSigma$ is the identity matrix.}

\begin{adjustbox}{width=\textwidth}
\begin{tabular}{r*{ 12}{r}}
\toprule
& & &  \multicolumn{3}{c}{$d$ = 100} & \multicolumn{3}{c}{$d$ = 200}  & \multicolumn{3}{c}{$d$ = 300} \\
 \cmidrule{4-12}
Model  & ($\rho$, $\sigma$) &  Method & TPR &  FPR & Time(s) & TPR &  FPR & Time(s) & TPR &  FPR & Time(s)  \\
 \midrule
9 & (0,\ 1.0) & ADMM  & 100.0\% & 0.33\%  & 0.28(0.05)   & 100.0\% & 0.23\% & 1.32(0.20) & 98.4\% & 0.10\% & 2.46(0.33)\\
& & IP & 99.2\% & 0.26\% & 0.07(0.01) & 96.6\% & 0.21\% & 0.15(0.02) & 91.4\% & 0.06\% & 0.31(0.04)\\
& & RAMP  & 0.84\% & 0.02\%  & 0.30(0.04)   & 2.01\% & 0.01\% & 1.41(0.23)  & 1.12\% & 0.00\%
& 2.53(0.36) \\
 & (0,\ 1.5) & ADMM  & 98.4\% & 0.25\%  & 0.37(0.04)  &  96.7\% & 0.23\%  & 1.44(0.26)
 & 95.2\% & 0.09\% & 2.60(0.37)  \\
& & IP  & 96.5\% & 0.22\% & 0.07(0.01) & 95.3\% & 0.13\% & 0.25(0.03) & 93.8\% & 0.08\% & 0.30(0.04)\\
& & RAMP   & 1.39\% & 0.01\% & 0.34(0.05)    & 1.69\% & 0.03\%  & 1.58(0.41) & 0.96\% & 0.00\%  & 2.43(0.49)\\
& (0.2,\ 1.0) & ADMM  & 99.9\% & 0.18\% & 0.35(0.04) & 99.7\% & 0.09\%  & 1.34(0.21) & 98.2\% & 0.03\%   & 2.61(0.40)\\
& & IP & 98.6\% & 0.13\%   & 0.06(0.02)  & 98.3\%  & 0.05\% & 0.16(0.02) & 95.0\% & 0.02\%  & 0.38(0.04)\\
& & RAMP  & 1.08\% & 0.00\% & 0.46(0.04)   & 1.35\% & 0.00\%  & 1.49(0.31) & 1.29\% & 0.00\% & 2.51(0.35)\\
 & (0.2,\ 1.5) & ADMM  & 98.9\% & 0.21\%   & 0.42(0.05)   & 96.9\% & 0.14\%  & 1.40(0.22) & 95.8\% & 0.07\%  & 2.59(0.43)\\
& & IP & 96.1\% & 0.24\%  & 0.08(0.01) & 94.2\% &  0.10\%  & 0.13(0.04)& 90.3\% & 0.03\% & 0.27(0.05)\\
& & RAMP  & 0.82\% & 0.00\% & 0.49(0.05) & 0.79\% & 0.00\% & 1.44(0.30)  & 0.66\% & 0.00\%  & 2.43(0.39)\\
  \bottomrule
 \end{tabular}
   \end{adjustbox}
 \label{tab:5}
 \end{table}

 \begin{table}
\centering
\caption{The means and standard errors (in parentheses) of testing errors and median model sizes in GPL96 data analysis for breast cancer.}

\vspace{0.5cm}

\begin{tabular}{{r}{c}{c}}
\toprule
Method & Classification Error & Median Model Size\\
\midrule
ADMM & 6.13\% (0.24\%) & 97\\
IP      & 7.38\% (0.22\%)  & 85\\
RAMP & 8.15\% (0.35\%) & 68\\
  \bottomrule
 \end{tabular}
 \label{tab:r1}
 \end{table}

 \begin{table}
\centering
\caption{The means and standard errors (in parentheses) of testing errors and median model sizes in GPL96 data analysis.}

\vspace{0.5cm}

{\footnotesize
\begin{tabular}{c*{6}{c}}
\toprule
 \multicolumn{2}{c}{ADMM} & \multicolumn{2}{c}{IP}  & \multicolumn{2}{c}{RAMP} \\
 \midrule
Interaction & Frequency & Interaction & Frequency & Interaction & Frequency \\
 \midrule
BRCA1 $\times$ LMO3 & 75 &  c-Myc $\times$ KLK3 & 69 & ESR1 $\times$ IRF4 & 64 \\
BRCA1 $\times$ PALB2 & 71 & FOXA1 $\times$ NF$\kappa$B & 62 &  SERINC5 $\times$ CHPF & 61 \\
 TP53 $\times$ RXRA & 64 &  JUND $\times$ GSTP1 & 56  &  EPCAM $\times$ RBP1  & 55\\
 BRCA2 $\times$ HNF4A & 59 & BRCA1 $\times$  ALCAM & 52 & MAF $\times$ LMOD1 & 51\\
 PDLIM5 $\times$ SETDB1 & 58  & DPT $\times$ SETDB1 & 51 &  BRCA2 $\times$ TARP & 51 \\
  \bottomrule
 \end{tabular}
}
 \label{tab:r2}
 \end{table}

   \begin{table}
\centering
\caption{The means and standard errors (in parentheses) of testing errors and median model sizes in GPL96 data analysis for Ewing Sarcomar.}
\vspace{0.5cm}
 \begin{tabular}{{r}{c}{c}}
\toprule
Method & Classification Error & Median Model Size\\
\midrule
ADMM & 5.58\% (1.81\%) & 79\\
IP      & 8.90\% (1.69\%)  & 71\\
RAMP & 19.72\% (4.55\%) & 95\\
  \bottomrule
 \end{tabular}
 \label{tab:re}
 \end{table}


\end{document}